# KINEMATICS AND DYNAMICS OF THE GALACTIC DISC

## Martín LÓPEZ-CORREDOIRA[1,2]

[1] *Instituto de Astrofísica de Canarias, E-38205 La Laguna, Tenerife, Spain*
*(email: fuego.templado@gmail.com)*
[2] *Departamento de Astrofísica, Universidad de La Laguna, E-38206 La Laguna, Tenerife, Spain*

**Abstract:** The Milky Way disc presents a warp, a flare, lopsidedness and other deviations from a purely axisymmetrical double exponential density component, both for the stellar and the gas component. Moreover, recent large-scale extended kinematics maps of Gaia data have shown significant departures from circularity in the mean orbits with radial Galactocentric velocities between -20 and +20 km/s and vertical velocities between -10 and +10 km/s, asymmetries northern/southern and other anomalies.

The observed features are sensitive to internal and external forces, especially at the largest Galactocentric distances. Here we review some of the Galactic dynamics hypotheses that explain these anomalies: either in terms of gravitational interaction, magnetic fields, accretion of intergalactic matter or others. The gravitational interaction may be among the different components of the Galaxy or between the Milky Way and another companion galaxy. The accretion of intergalactic matter may be either into the halo, with a later gravitational interaction between the misaligned halo and the disc, or directly onto the disc.

Most of the relevant observations are explained in terms of the accretion of intergalactic matter onto a disc that is far from a simple stationary configuration in rotational equilibrium. Other hypotheses only partially explain some observations.

## 1. The Galactic disc

We live in a spiral galaxy, which has a prominent disc component in which the spiral arms are embedded, together with other parts, such as the halo and bulge/bar.

The surface brightness distribution in spiral galaxies is usually an exponential law in the radial direction (de Vaucouleurs 1959; Freeman 1970), i.e. *$I(R)=I_0$ exp$(-R/h_R)$*, $h_R$ being the scale length. This law is also followed in our Galaxy (de Vaucouleurs & Pence 1978). Moreover, in the vertical direction perpendicular to the plane defined by the disc, another approximately exponential dependence is observed, so globally the density follows a dependence *$\rho(R,z)= \rho_0$ exp$(-R/h_R)$ exp$[-|z|/h_z(M)]$* (Bahcall & Soneira 1980), with a scale height $h_z$ that depends on the population. Historically, however, there also other attempts to fit the vertical distribution with an isothermal vertical *$sech^2(z)$* distribution (Villumsen 1983), which is used less frequently nowadays owing to its poorer fits.

As in most spiral galaxies (Comerón et al. 2011), the disc of the Milky Way is indeed a double disc: with a dominant thin disc with scaleheight $h_z$~0.3 kpc for the dwarf population of the average disc (Bahcall & Soneira 1980) and a thick disc with only very old population and $h_z$~*1* kpc (Gilmore & Reid 1983). Since the 1980s, many hundreds of papers have been produced with many different values of the scale heights of the two components, but the orders of magnitude are still the same. The distinction between the two discs is appreciated not only in the density distribution but also in [α/Fe] vs. [Fe/H] diagrams (Casagrande et el. 2011; Haywood et al. 2013). There are also gradients in the metallicity distribution in both discs in the radial (Xiang et al. 2015; Tunçel et al. 2019) and vertical directions (Ak et al. 2007; Tunçel et al. 2019).

Some spiral galaxies present a gradual density decrease (e.g. type II truncation; Freeman 1970) in the inner disc. Many barred galaxies even have holes (Ohta et al. 1990). The Milky Way inner disc also presents a deficit of CO gas (Robinson et al. 1988), total near-infrared flux (Freudenreich 1998) and stellar density (López-Corredoira et al. 2004). However, there are no signs of truncation in the outer disc. Some authors (e.g., Minniti et al. 2011; Amôres et al. 2017) have pointed out there is a cut-off or lower than the value of the extrapolation of exponential density at $R>13$-16 kpc, but they did not take into account the flare (see next paragraph). The disc stars (distinguished from halo stars by their high metallicity) do indeed extend out to at least $R \approx 25$ kpc (López-Corredoira et al. 2018), with an exponential distribution (López-Corredoira & Molgó 2014); and the gas extends in an exponential distribution out to at least $R \approx 40$ kpc (Kalberla & Dedes 2008). Non-axisymmetrical density distributions in discs, called 'lopsidedness', is observed in ~30% of the galaxies (Baldwin et al. 1980; Zaritsky & Rix 1997; Jog & Combes 2009). The Milky Way gas presents some degree of this type of asymmetry: the large scale HI distribution is lopsided at $R>20$ kpc (Kalberla & Dedes 2008).

The scale height of the disc depends on both the Galactocentric distance $R$ (Narayan & Jog 2002; Nakanishi & Sofue 2003) ('flare'); and on the azimuth (Voskes 1999; Levine et al. 2006; Kalberla et al. 2007). And there is a flare in the stellar distribution too (López-Corredoira et al. 2002a; López-Corredoira & Molgó 2014; Wang et al. 2018), in both the thin and thick discs (Mateu et al. 2011; López-Corredoira & Molgó 2014).

The warp is the other most prominent feature in the outer disc of our Galaxy, as well as in most spiral galaxies (Sánchez-Saavedra et al. 2003). In the gas distribution, a warp has been well known for a long time (Burke 1957; Voskes 1999; Levine et al. 2006; Kalberla & Dedes 2008). The warp in stars distribution was discovered more recently, both for the old population (López-Corredoira et al. 2002a; Reylé et al. 2009) and young stars (Amôres et al. 2017; Skowron et al. 2019). An asymmetry between the northern and southern warp at $R>15$ kpc, with a higher amplitude in the north than in the south, is observed both for the gas (Voskes 1999) and the young population (Amôres et al. 2017).

A discussion of the dynamical reasons for these observed phenomena is given in Sections 3–5.

## 2. Extended kinematic maps of Gaia-DR2

The large-scale kinematics of the disc has been explored more recently. Much literature has been produced in recent decades, but we focus here on an analysis with data from the Gaia mission, which surpasses many previous achievements. The major exploitation of 6D phase space (3D spatial + 3D velocity) maps has indeed arrived with the kinematic maps of Gaia data (Gaia Collaboration 2018), the disc being the component with better prospects of analysis owing to the low extinction of the sources around the Sun and towards the anticentre.

Gaia Collaboration (2018) analyses the whole sample of stars from Gaia-DR2 including radial velocity measurements, which consist of more than seven million sources. They have provided kinematic maps within 5 kpc of the Sun, which is a reachable range for stars with relative error in distance lower than 20%. Applying a statistical deconvolution of the parallax errors based on the Lucy inversion method of the Fredholm integral equations of the first kind without assuming any prior, López-Corredoira & Sylos Labini (2019) have extended the maps in a range of heliocentric distances by a factor of two to three larger with respect to Gaia Collaboration (2018), up to $R$=20 kpc. This extension to farther distances is interesting for the kinematical studies of the disc because many relevant features away from an axisymmetric disc in equilibrium occur at $R$>13 kpc. The warp, flare, and most significant fluctuations with respect to zero radial or vertical velocities, take place where the density of the disc is lower, so it is worth extending the analysis beyond 13 kpc from the Galactic centre.

The extended maps of López-Corredoira & Sylos Labini (2019) provide a lot of new and corroborated information about disc kinematics. One of the main results is that the radial component of the velocity displays considerable gradients. In particular Fig. 8/top of López-Corredoira & Sylos Labini (2019) shows significant radial Galactocentric velocities ($V_R$) between -20 and +20 km/s. The expansion/contraction affects different parts of the disc differently: in their Fig. 8/top-left, we see expansion ($V_R$>0) for azimuths -5º<$\Phi$< 60º and $R$>10 kpc, whereas there is contraction ($V_R$<0) for azimuths -50º<$\Phi$< -5º and $R$>10 kpc, although only within |$\Phi$|<~25º the errors are lower than 10 km/s, which makes the detection significant. Large scale motions corresponding to expansion and contraction in different parts of the outer disc are taking place, which suggests giant processes of inflows and outflows of stars are occurring in our Galaxy. The observed vertical velocities between -10 and +10 km/s with some correlated gradients also require the existence of vertical forces in Galactic dynamics. Other features are observed, such as variations in azimuthal velocity with position; or asymmetries between the northern and the southern Galactic hemispheres, especially towards the anticentre, which includes a larger azimuthal velocity in the south. The Jeans equation relates azimuthal velocity and rotation speed, so azimuthal velocity data are used to derive the rotation speed (Chrobáková et al. 2019), with the astonishing result that it is almost independent of the vertical distance from the plane ($z$).

A discussion of the dynamical reasons for these observed phenomena is given in Section 6.

### 3. Dynamics: Double exponential thin+thick discs

Explanations for the stability and the exponential radial distribution have been proposed for many decades. One of the first suggestions stems from ideas of disc galaxies formation, according to which extended dark matter haloes are formed first, the thick disc and bulge later, and finally the exponential thin disc from gas accretion (Fall & Efstathiou 1980; Jones & Wyse 1983). In this process of disc formation, viscous redistribution of gas is essential for ensuring an exponential disc (Ferguson & Clarke 2001). If the viscous time scale equals the time scale of star formation, the disc is exponential independently of the rotation curve and viscosity prescription (Lin & Pringle 1987; Yoshii & Sommer-Larsen 1989). Exponential and quasi-exponential discs in the radial direction can also be produced through a combination of supernova-driven galactic outflows, intrinsic variation in the angular momentum distribution of the halo gas and the inefficiency of star formation at large radii (Dutton 2009).

For the vertical distribution, first theoretical attempts to model it pointed out that it should be an isothermal distribution with sech$^2$ law (Spitzer 1942). This is derived from joining the hydrostatic equilibrium equation and the Poisson equation, assuming negligible radial acceleration (in a thin disc) and azimuthal acceleration (in an axisymmetrical disc). As a result of the combination of these two hypotheses, we get $\rho(z)=A\ \mathrm{sech}^2[a(z-z_0)]$. But, as described in section 1, we know that the vertical distribution is not a sech$^2$ law, but it is closer to an exponential distribution, it is steeper in the regions of low *z*. To explain those exponential *z*-profiles, Burkert & Yoshii (1986) use models of gaseous protodiscs that settle into isothermal equilibrium prior to star formation and subsequent gas cooling at a rate comparable to the star formation rate. Another proposal was given by Banerjee & Jog (2007), with gas concentrated closer to the Galactic mid-plane (because of its low velocity dispersion) than the stars; hence, it strongly affects gravitationally the vertical stellar distribution.

The existence of a thick disc separate from the thin disc is usually understood as a vertical dispersion increase with radius and age due to heating of the disc. This heating may be due to collision with accreted gas, satellites, heating, or radial migration (Sales et al. 2009; Qu et al. 2011; Haywood et al. 2013).

### 4. Dynamics: Flare, non-axisymmetrical scale height and lopsidedness

The isothermal solution that we introduced in the last section, even with non-negligible radial acceleration (allowing for a thick disc), gives a very small flare, not enough to explain the observed increase of scale height with Galactocentric distance. There are other ideas in the literature that try to explain this phenomenon. First of all, one must consider the gravitational coupling between disc stars, disc gas and the dark matter halo (Narayan & Jog 2002), and this explains the flare well. Apart from self-gravity, other effects may contribute in some degree, such as the presence of magnetic fields (Battaner & Florido 1995; note however that Sánchez-Salcedo & Reyes-Ruiz 2004 claim that there was an error in the calculations of Battaner & Florido), or the pressure due to accretion of intergalactic matter (López-Corredoira & Betancort-Rijo 2009). The non-axisymmetrical scale height, by the way, is also explained by this last hypothesis, as well as the lopsidedness (Bournaud et al. 2005). Kalberla et al. (2007) suggested a more exotic proposal for this dependence on azimuth: a ring of dark matter embedded in the galactic disc with a radius that depends on the azimuth.

There are other proposals to explain lopsidedness. Baldwin et al. (1980) suggested a lopsided pattern of elliptical orbits in which all the lines of apsides rotate at the same rate. Dynamical instability due to spontaneous growth of modes which are long-lived could be another explanation (Saha et al. 2007; Dury et al. 2008). Also, the response to a distorted halo (Jog 1997) or disc lying off-centre in the potential of the halo (Levine & Sparke 1998). Mergers, tidal interaction and, as mentioned, gas accreted onto the disc (Bournaud et al. 2005) are also possible explanations.

### 5. Dynamics: Warp

Warps are observed in most spiral galaxies (Sánchez Saavedra et al. 2003), so they must be either formed very frequently during the life of a galaxy, or, if they are formed only once, they should be maintained long term. The idea of warps formed and maintained during the entire life of an isolated spiral galaxy without interaction with external forces was explored a short time after their discovery and rejected by Hunter

& Toomre (1969). Nonetheless, it has been more recently revived: some hypotheses about warps in isolated discs have been developed, such as bending wave instabilities or vertical resonance caused by a centrifugal force of particles following a bend in a disc, possibly due to the bar, which is higher than the gravitational restoring force of the disc itself (Griv et al. 2002; Revaz & Pfenninger 2004; Sánchez-Martín et al. 2016; Chequers & Widrow 2017). Such mechanisms, however, generate low amplitude S-warps, and it is not clear that they can live long enough.

The most accepted idea is that a warped galactic disc can be modelled as a set of rotating rings (Rogstad1974), and an external torque exerted on each ring of the outer disc would produce the rotations of the rings that constitute the warp. The external torques may be due to tidal gravitational interactions with another galaxy (Elbert & Hablick 1965; Huang & Carlberg 1997), or its amplification through a massive halo (Weinberg 1998), extragalactic magnetic fields (Battaner et al. 1990, 1991), misalignment of very-massive halos due to accretion of intergalactic matter into the halo (Ostriker & Binney 1989; Jiang & Binney 1999) or in its formation (Dubinski & Chakrabarty 2009; Roškar et al. 2010), or accretion of intergalactic matter onto the galactic disc (Mayor & Vigroux 1981; Revaz & Pfenninger 2001; López-Corredoira et al. 2002b; Sánchez-Salcedo 2006; Haan & Braun 2014), or impulsive encounters with satellites (Kim et al. 2014).

The tidal gravitational interaction with the Magellanic Clouds or the Sagittarius dwarf galaxy is insufficient to produce the warp of the Milky Way (Hunter & Toomre 1969; Bailin 2003). Even with the amplification of Magellanic Clouds tides by the halo (Weinberg 1998), the model does not fit for the Magellanic Clouds (García-Ruiz et al. 2002a).

An intergalactic magnetic field of amplitude 1 μG in interaction with the Galactic magnetic field could produce a torque leading to a warp (Battaner et al. 1990; Battaner & Jiménez-Vicente 1998). This is an interesting and plausible idea, but regrettably, the community, obsessed with dark matter haloes, has not paid enough attention to magnetic fields solutions.

The gravitational torque produced by misaligned haloes (Ostriker & Binney 1989; Jiang & Binney 1999; Dubinski & Chakrabarty 2009; Roškar et al. 2010) is perhaps the most extended idea. However, if there is a continuous infall of intergalactic matter into the halo, it is not clear how the very low density of baryonic matter in the halo can trap the infalling matter.

One may better understand the accretion when it is onto the galactic disc. Integral-shaped (S-shaped) warps as a consequence of the transmission of angular momentum are produced in Milky Way type galaxies with an infall velocity of ~100 km/s, an intergalactic baryon density of ~$10^{-25}$ kg/m$^3$, leading to an accretion of ~1 solar mass per year (López-Corredoira et al. 2002b), compatible with our knowledge about the ranges of these numbers. The global angular momentum of the intergalactic medium is null, but the redistribution of momentum in different rings of the galaxy due to gravity produces a net torque in each ring. Moreover, this hypothesis can explain the much less frequent cup-shaped (U-shaped) warps due to a transmission of linear momentum that is not explained by previous theories. Asymmetric cases, such as the Milky Way warp, can also be explained by the present theory as a combination of S- and U-warps (Saha & Jog 2006). Also, in this scenario, we can understand why the frequency of warps and their amplitude are dependent on the environment, as observed (García-Ruiz et al.

2002b). And we also understand with this hypothesis the coincidence of the same angles of northern-southern warp ($\Phi$=90 or 270 deg.) with the maxima or minima or $h_z(\Phi)$ (López-Corredoira & Betancort-Rijo 2009).

Therefore, there are good reasons for preferring accretion onto the disc rather than into the halo. As mentioned, the halo density of baryonic matter is very low to trap intergalactic matter, but the disc density is not. There are also observations that indicate that the amplitude of the warps is not correlated with the dark matter ratio (from rotation curves; Castro-Rodríguez et al. 2002), whereas some correlation would be expected if the role of the halo were important. The gas is more warped than stars in the Milky Way (Reylé et al. 2009), and no lenticular galaxy is warped (Sánchez-Saavedra et al. 2003), so gas seems to be necessary in the disc to form warps rather than a purely gravitational halo-disc interaction that would affect stars and gas equally. Moreover, other facts point to the necessity of accretion of intergalactic matter onto the disc (Sánchez Almeida et al. 2014).

## 6. Dynamics: Kinematic features

The connection between kinematics and dynamics has been intensively studied in recent decades of research into the Milky Way as a galaxy. Here, we simply describe some achievements based on the analysis of the large-scale extended kinematic maps of Gaia DR2 (see Section 3). A compendium of different dynamical mechanisms whose predictions can be compared to these kinematic maps is given by López-Corredoira et al. (2019) for radial and vertical velocities, and by Chrobáková et al. (2019) for the azimuthal velocities.

The conclusion of these studies is that many mechanisms may generate either non-null radial velocities, or non-null vertical velocities. The gravitational influence of components of the Galaxy other than the disc, such as the bar/bulge or spiral arms or tidal interaction with Sagittarius dwarf galaxy, may explain some features of the velocity maps, especially in the inner parts of the disk. However, they are not enough to explain the most conspicuous gradients in the outer disk.

Vertical motions might be dominated by external perturbations or mergers, although with a minor component due to the warp. López-Corredoira et al. (2019) carried out *N* body simulations to investigate the possible contribution of the minor merger to the vertical asymmetrical bulk motions; they found that a minor merger with some satellites could explain the positive vertical velocity on both north and south side of the hemisphere in the outer disc.

There are two contributions of the warp to the vertical motion: one produced by the inclination of the orbits and another from the variation of the amplitude of the warp angle, which is significantly detected in the fits to the data, the most likely period of the oscillation of the warp being around 0.5 Gyr, although much longer periods are not excluded (López-Corredoira et al. 2019). As said in Section 5, transient warps may be related to a variable torque over the disc; for instance, when the torque is produced by the misalignment of halo and disk, and when the realignment is produced in less than 1 Gyr (Jiang & Binney 1999), or in a scenario of accretion of intergalactic matter with variable ratios of accretion (López-Corredoira et al. 2002b).

Kinematic distributions, including information on the dispersion of velocities, can also be related to the width of the disc. López-Corredoira et al. (2019), with two different methods using Jeans equation, have seen that the mass distribution of the disc is flared, i.e. the thickness of the disc increases outwards. The numbers obtained are roughly consistent with previous analyses of the flare based only on the morphology (López-Corredoira & Molgó 2014, and references therein), so we can connect both the increasing scale height and dispersion of velocities outwards as the same phenomenon. Nonetheless, we must also consider that non-equilibrium systems do not strictly follow the Jeans equation. Haines et al. (2019) show that traditional Jeans modelling should give reliable results in overdense regions of the disc, but important biases in underdense regions, calling for the development of non-equilibrium methods to estimate the dynamical matter density locally and in the outer disc. Further analyses of this deviation of the Jeans equation in non-equilibrium systems for the application of the present Gaia data are explored by Chrobáková et al. (2019).

Precisely, the non-equilibrium system is one conclusion. In the lack of other possible causes for the main observed features in the kinematic maps, they can only find an explanation in terms of models in which the Galactic disc is still in evolution, either because the disc has not reached equilibrium since its creation or because external forces (such as the Sagittarius dwarf galaxy) might perturb it. López-Corredoira et al. (2019) have explored a simple class of out-of-equilibrium, rotating, and asymmetrical mass distributions that evolve under their own gravity. Non-circular orbits and with significant vertical velocities in the outer disc are precisely one of the predictions of this model. Orbits in the very outer disc are out of equilibrium so they have not yet reached circularity, precisely as we observe in our data. The large velocity gradients observed in the Gaia DR2 data are at odds with the simple model in which stars move on steady circular orbits around the centre of the Galaxy. The discussed non-equilibrum model provides a qualitative framework in which such non-stationarity is intrinsic to the dynamical history of the Galaxy, rather than being induced by an ad hoc external field due to the passage of a satellite galaxy.

Rotation curves are another basic ingredient of the kinematics of the Milky Way. Chrobáková et al. (2019) show similar curves in in-plane and off-plane regions for the stellar component. In the plane, the results are similar to the gas kinematics (Sofue et al. 2009). As is well known, the fits of the rotation speed in the plane can be explained by both the existence of a very massive dark matter halo or modified gravity (e.g. MOND), but the fact that off-plane regions show very little decrease of velocity in principle favours the first hypothesis (Chrobáková et al. 2019). Nonetheless, we should bear in mind that the Jeans Equation requires equilibrium, and non-equilibrium effects may be significant (Haines et al. 2019), so rotation speeds in very low density regions must be treated with caution.

Apart from rotation speed analyses, the analysis of the dispersion of velocities through the Jeans equation is also noteworthy. It has led to the discussion over the last decade of the lack (Moni-Bidin et al. 2012, 2015) or presence (Bovy & Tremaine 2012; Loebman et al. 2012; Sánchez-Salcedo et al. 2016; Hagen & Herlmi 2018) of a dark matter halo. There is no general agreement on the question. In any case, we must note again that the Jeans equation application might be inappropriate because of non-equilibrium in low density regions (Haines et al. 2019; Chrobáková et al. 2019).

## 7. Summary of theories and discussion

The morphology of the Milky Way has been known through the many analyses of recent decades. Its large-scale kinematics became known only recently, especially with the big leap taken with the arrival of Gaia data, plus the rotation curves that have been known for a long time. The observed features are sensitive to internal and external forces, especially at the largest Galactocentric distances. Many hypotheses may partially explain them. Here we have covered the main ideas on the dynamics of the Milky Way. There are five main theories that may explain three or more of those features. Here we give a summary:

1. Gravitational interaction with a satellite: it explains lopsidedness, radial velocities, vertical velocities, and may explain the existence of the thick disc.
2. Minor mergers: it explains lopsidedness, radial velocities, vertical velocities, and may explain the existence of the thick disc.
3. A non-axisymmetrical component (bar, spiral arms, ring of dark matter): azimuthal dependence of scale height, radial velocities in the inner disc, holes/type-II-truncation in the inner disc.
4. Accretion of intergalactic matter or small satellites onto the disc: it explains the thick disc formation, lopsidedness, vertical velocities, flares, azimuthal dependence of scale height, S-warps, U-warps. Moreover, other facts besides the disc morphology or kinematics (chemical evolution, bar formation, spiral arms, maintaining spiral structure, etc.) fit within this scenario.
5. Non-equilibrium in an isolated Milky Way: it explains lopsidedness, radial velocities, vertical velocities.

Other elements may also be necessary to explain some particular observation.

Therefore, many possible views of the Milky Way are possible. A scenario of formation of discs with several components plus a dark matter massive halo, star formation efficiency, accretion of intergalactic matter onto the disc and non-equilibrium (either in an isolated Milky Way or due to interaction with satellites) might be enough to explain the most relevant features.

*Acknowledgements:* Thanks are given to T. J. Mahoney for proof-reading this text.